\documentclass[useAMS]{mn2e}
\usepackage[utf8]{inputenc}
\usepackage[dvips]{graphicx}

\title[Super-Earths rotation]
{Spin-orbit coupling for tidally evolving super-Earths}
\author[A. Rodr\'iguez et al.]{A.~Rodr\'iguez,$^1$N. Callegari Jr.$^2$, T. A.~Michtchenko,$^1$ and H. Hussmann$^3$\\
$^1$ Insituto de Astronomia, Geof\'isica e Ci\^encias Atmosf\'ericas, IAG-USP, Rua do Mat\~ao 1226, 05508-900, S\~ao Paulo, Brazil\\
$^2$ Instituto de Geoci\^encias e Ci\^encias Exatas, Unesp-Univ Estadual Paulista , Av. 24-A, 1515, 13506-900, Rio Claro, SP, Brazil\\
$^3$ German Aerospace Centre (DLR), Institute of Planetary Research, Rutherfordstr. 2, 12489, Berlin, Germany}

\date{Released 2012 Xxxxx XX}

\onecolumn
\begin{document}

\maketitle

\begin{abstract}

We investigate the spin behavior of close-in rocky planets and the implications for their orbital evolution. Considering that the planet rotation evolves under simultaneous actions of the torque due to the equatorial deformation and the tidal torque, both raised by the central star, we analyze the possibility of temporary captures in spin-orbit resonances. 

The results of the numerical simulations of the exact equations of motions indicate that, whenever the planet rotation is trapped in a resonant motion, the orbital decay and the eccentricity damping are faster than the ones in which the rotation follows the so-called pseudo-synchronization. Analytical results obtained through the averaged equations of the spin-orbit problem show a good agreement with the numerical simulations. 

We apply the analysis to the cases of the recently discovered hot super-Earths Kepler-10\,\textbf{b}, GJ 3634\,\textbf{b} and 55 Cnc\,\textbf{e}. The simulated dynamical history of these systems indicates the possibility of capture in several spin-orbit resonances; particularly, GJ 3634\,\textbf{b} and 55 Cnc\,\textbf{e} can currently evolve under a non-synchronous resonant motion for suitable values of the parameters. Moreover, 55 Cnc\,\textbf{e} may avoid a chaotic rotation behavior by evolving towards synchronization through successive temporary resonant trappings.

\end{abstract}


\section{Introduction}\label{intro}

Tidal theories have gained great attention in the last decade due to the large number of discovered close-in planets which orbit their parent stars with periods of a few days. According to classical theories (e.g.  Darwin 1880; Kaula 1964; Mignard 1979), the long-term tidal evolution of these planets results in orbital decay and circularization, on time-scales which depend on the physical properties of the interacting systems (Hut 1981; Dobbs-Dixon, Lin \& Mardling  2004; Ferraz-Mello, Rodr\'iguez \& Hussmann 2008; Jackson, Barnes \& Greenberg 2009; Rodr\'iguez \& Ferraz-Mello 2010).  

The rotation of short-period planets is also affected by tidal interactions. It is nowadays accepted that the spin rate of planets soon after their formation was much higher than their orbital mean motion (e.g. Agnor, Canup \& Levison 1999; Kokubo \& Ida 2007).  However, the cumulative tidal effects causes the planet to spin down, reaching its stationary value in time-scales much shorter than those needed to circularize the planet orbit. This evolution ultimately results in the synchronous rotation (the spin state in which the orbital and rotation periods coincide), which occurs when the planet orbit is circularized (Hut 1981; Ferraz-Mello et al. 2008). For non-circular orbits, the synchronous rotation only occurs if an additional torque acts to counteract the tidal torque.

The additional gravitational torque may be generated by planets with a rocky-like composition having a permanent equatorial deformation or frozen-in figure (see Goldreich \& Peale 1966; Greenberg \& Weidenschilling 1984). These torques, in contrast to the tidal ones, are conservative. In the context of the two-body problem, the gravitational interaction of the star with an asymmetric planet drive the planet rotation into different regimes of motion, including oscillations around exact spin-orbit resonances (Goldreich \& Peale 1966). When dissipative tidal effects are taken into account, the oscillations are damped and the planet rotation is trapped into a resonance. Indeed, the rotational motion of some Solar System bodies can be explained by the combined effects of the tidal and asymmetric torques (Goldreich \& Peale 1966; Greenberg \& Weidenschilling 1984; Correia \& Laskar 2009). 

Mercury is an example of a rocky planet with a rotation state currently evolving in the 3/2 spin-orbit resonance (Pettengill \& Dyce  1965).   Several models of the secular spin evolution of Mercury have been developed, including i) the chaotic evolution of its orbit (Correia \& Laskar 2004); ii) effects of the core-mantle friction (Correia 2006; Correia \& Laskar 2009); iii) the evolution starting with retrograde and subsequent synchronous motion combined with impact events (Correia \& Laskar, 2012; Wieczorek et al. 2012); iv) the consideration of non-linear tidal theories (Makarov 2012). However, despite the large amount of previous works, the question on how the current resonant state of Mercury was achieved still remains under discussion.

Concerning the rotation of Earth-like close-in extrasolar planets, a near synchronous motion is usually assumed in view of tidal evolution. However, Correia, Levrard \& Laskar (2008) have shown that when an additional torque is considered, the equilibrium rotation of tidally evolved planets can assume rotational states which depart from synchronism. Moreover, as the capture probability in a non-synchronous spin-orbit resonance requires a non-circular orbit (Goldreich \& Peale 1966; Celletti \& Chierchia 2008), the rotation of several recently discovered super-Earth planets in eccentric orbits can be currently evolving under a resonant motion, provided the planet figure can support a permanent equatorial deformation. As the eccentricity is tidally damped, several resonant trappings would become unstable and the planet rotation ultimately reaches synchronization for almost circular orbits. However, the locking in a non-synchronous resonance with the very short orbital period can protect certain planet’s longitudes from solar radiation, providing more favorable conditions with respect to habitability as compared to non-resonant rotation (Dobrovolskis 2007). Moreover, Heller, Leconte \& Barnes (2011) studied the equilibrium rotation of super-Earths and the implications on their habitability. They found that, under assumption of an evolution driven by tidal torques, the equilibrium value of the rotation period of GJ 581 \textbf{d} is probably about two times smaller than the orbital period. 

In a recent work, Callegari \& Rodr\'iguez (2012) have analysed the rotation of extrasolar planets with orbital periods less than 33 days and masses below the mass of Neptune. Depending on one parameter related to the equatorial ellipticity and also on the orbital eccentricity, they identified regions of chaotic rotation in the phase space of some particular planets like 55 Cnc\,\textbf{e}, GJ 1214\,\textbf{b}, Gliese 876\,\textbf{d}, among others.

The present investigation aims at two main aspects: i) How the orbital evolution of the planet, due to tidal interaction with the central star, is altered when the asymmetric torque is considered; and ii) To analyse the occurrence of captures in spin-orbit resonant motions under combined effects of both torques for the planets Kepler-10\,\textbf{b}, 55 Cnc\,\textbf{e} and GJ 3634\,\textbf{b}. These planets are examples of hot super-Earths, which are close-in planets having a few Earth-masses with a rocky-like composition. We are interested in the past behavior in the history of the rotation and the subsequent evolution to synchronization. We also discuss the possibility of a current evolution of the rotation of these planets into a super-synchronous resonance state. Moreover, we investigate the implications on the orbital evolution due to such temporary resonant trappings. 

Despite the continuously growing number of discovered planets, their rotation state and equilibrium figures remain unknown. In this work, we perform several numerical simulations of the evolution of the systems, assuming several values of unknown parameters, such as equatorial ellipticity and initial rotation. The results allow us to understand the past history of these systems, their current orbital configuration and final rotational states. It should be noted that the majority of previous works concerning spin-orbit interactions with tidal dissipation have been addressed in the context of a fixed Keplerian orbit, with constant values of semi-major axis and eccentricity (e.g., Celletti, Froeschl\'e \& Lega 2007; Celletti \& Chierchia 2008).  This assumption is adequate for Solar System bodies, where tidal interactions are weak and, consequently, variations of orbital elements are small.  

On the other hand, the existence of extrasolar planets with orbital periods of a few days indicates that their current orbital configurations would be the consequence of strong tidal interaction with the host star. In this case, the Keplerian approximation is no longer valid. Thus, we analyse spin-orbit interactions as the orbit evolves due to tidal effects. Solving the exact equations of motion of the problem, we show that the forces arising from the potential of a asymmetric planet also contributes to the orbital variation, accelerating the orbital decay and the eccentricity damping, but only in the cases in which the rotation is locked in resonance. 

The spin-orbit problem can also be modeled by the use of averaged equations of motion, where only terms with secular and resonant arguments are retained (e.g. Celletti \& Chierchia 2008; Correia \& Laskar 2009). Through this model, we compute the averaged variations of semi-major axis and eccentricity due to the combination of asymmetric and tidal forces. We also give an explicit expression for the critical value of the equatorial prolateness, for which a given spin-orbit resonance would become unstable. The study of the particular cases of 2/1, 3/2 and 1/1 spin-orbit resonances indicates a good agreement with the results of the numerical simulations of the exact equations.

This paper is structured as follows. In Section \ref{model}, we present the exact equations of motion of the spin-orbit problem. Section \ref{num} shows the results of the numerical simulations, illustrating the evolutions of Kepler-10\,\textbf{b}, GJ 3634\,\textbf{b} and Cnc 55\,\textbf{e}. We show that, assuming a nonzero value of the equatorial ellipticity, different temporary resonant trappings of the rotation can occur, with implications in the orbital decay and circularization. An analytical approach in the context of the averaged equations is shown in Section \ref{mean}, explaining the results obtained in the numerical simulations. Finally, Section \ref{discussion} is devoted to discussion and conclusions.

\section{The model}\label{model}


\begin{figure}
\begin{center}
\includegraphics[height=0.3\columnwidth,angle=0]{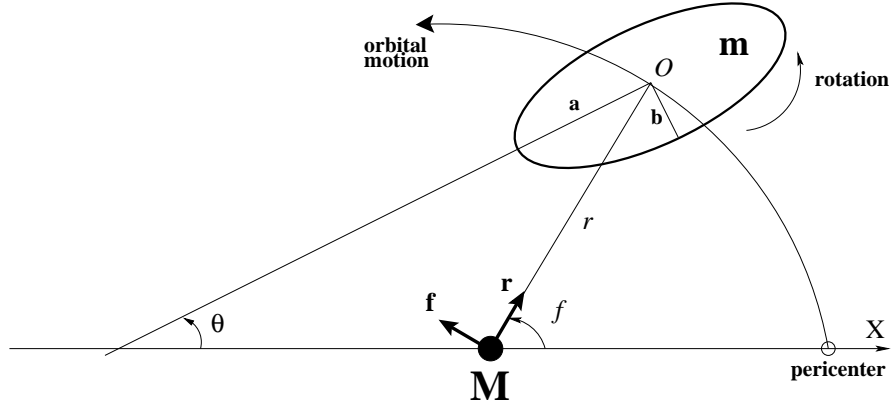}
\caption{\small Schematic view of the ellipsoid, showing the angles used in the formulation of the problem. The long axis makes an angle $\theta$ with the reference axis $X$, which is assumed coincident with the orbital pericenter direction . The angle between the long axis and the star-planet direction is given by $f-\theta$, where $f$ is the true anomaly of the planet.}	
\label{elips}
\end{center}
\end{figure}

We consider a planet orbiting a central star with a period of a few days. Hereafter, we refer to $m$ and $M$  as the planet and star masses, respectively; $R$ and $r$ as the planet radius and the instantaneous star-planet distance, respectively. The elements $a,e,n$ stand for the semi-major axis, eccentricity and mean-motion of the astrocentric orbit of the planet, respectively.

We suppose that the rotating planet body can be approximated by a homogeneous triaxial  ellipsoid, with the equatorial axes $\mathbf{a}$ and $\mathbf{b}$ and the rotation axis $\mathbf{c}$. We assume that $\mathbf{a}>\mathbf{b}>\mathbf{c}$ and  $\mathbf{c}$ is normal to the orbital plane (i.e., no obliquity is considered; planar problem). In addition, we assume that the equatorial asymmetry ($\mathbf{a}\neq\mathbf{b}$) corresponds to a frozen-in body or permanent deformation of the planet. The assumption above can be supported under consideration of a rocky composition of the planet body. This is the case of super-Earth planets, which are the subject of the present investigation. The gravitational potential of the planet at the distance $r$ from its centre is given by (e.g. Beutler 2005, p. 102)
\begin{equation}\label{u22}
U_{22}=3\frac{GmR^2}{r^3}C_{22}\cos2(f-\theta),
\end{equation}
where $G$ is the gravitational constant and the angles\footnote{We choose the pericenter of the planet orbit as the origin for the calculation of the longitudes because the main contribution to the variation of the longitude of the pericenter comes from the zonal term proportional to $J_2$, which has not been included in the present paper (see Ferraz-Mello et al. 2008).} $f$ and $\theta$ are defined in Figure \ref{elips}. Note that $\dot{\theta}=\Omega$ is the angular rotation velocity of the planet. The parameter $C_{22}$ describes the equatorial ellipticity of the planet's gravity field and is related to the principal momenta of inertia $A,B,C$ through $C_{22}=(\xi/4)(B-A)/C$, where $\xi$ is the structure constant given by $\xi=C/(mR^2)$ and $C$ is the polar moment of inertia.

We neglect the contribution of terms arising from the oblateness of the rotating body, that is, $J_2$ and higher order terms. In fact, in the planar problem, the gravitational potential corresponding to the $J_2$--term depends only on the distance $r$. In this case, the potential generates a radial force with zero torque that does not affect the rotation evolution.

At the distance $r$ from the centre of the planet, the gravitational field $U_{22}$ gives rise to a force on a point mass $M$ which is given by\footnote{Since we use the expression for $U_{22}$ adopted in Beutler (2005), the corresponding force is computed as the positive gradient of the potential.}
\begin{equation}\label{f22}
\mathbf{F}_{22}=M\,\nabla U_{22}=M\frac{\partial U_{22}}{\partial r}\,\mathbf{\hat{r}}+\frac{M}{r}\frac{\partial U_{22}}{\partial f}\,\mathbf{\hat{f}},
\end{equation}
where $\mathbf{\hat{r}}$ and $\mathbf{\hat{f}}$ are unit vectors with origin in the star (see Fig. \ref{elips}). The force $\mathbf{F}_{22}$ alters the orbital component of the angular momentum, whereas the reaction force governs the evolution of the spin through the creation of the torque on the planet given by $\mathbf{T}_{22}=-\mathbf{r}\times\mathbf{F}_{22}=-M(\partial U_{22}/\partial f)\,\hat{\mathbf{k}},$
where $\mathbf{r}=r\,\mathbf{\hat{r}}$ and $\hat{\mathbf{k}}=\mathbf{\hat{r}}\times\mathbf{\hat{f}}$. Thus,
\begin{equation}\label{t22}
\mathbf{T}_{22}=6\,\frac{GmMR^2}{r^3}C_{22}\sin2(f-\theta)\hat{\mathbf{k}}.
\end{equation}
Note that, for $C\,\dot{\Omega}=T_{22}$ and $C_{22}=(\xi/4)(B-A)/C$, the above equation agrees with the classical expression for the torque (e.g. Danby 1962; Goldreich \& Peale 1966).

The effects of the dynamical tides are raised by the gravitational action of the star on the inelastically deformed rotating planet. Due to the internal viscosity of the body, there exists a delay $\Delta t$ (time lag) between the star perturbation and the maximum deformation. The relationship between $\Delta t$ and tidal frequencies (which are linear combinations of $\Omega$ and $n$) depends on the internal rheology, which is poorly known even for Solar System bodies. For sake of simplicity, in this work, we adopt a linear model with constant $\Delta t$ (Darwin 1879; Mignard 1979; Hut 1981; Rodr\'iguez et al. 2011a); for a review of other tidal models, the reader is referred to Efroimsky \& Williams (2009) and Ferraz-Mello (2012).

According to Mignard (1979), the tidal force and torque are given by
\begin{equation}\label{mignardF}
\mathbf{F}_{{\textrm{\scriptsize tide}}}=-3k_{2}\Delta t\frac{GM^2R^5}{r^{10}}[2\mathbf{r}(\mathbf{r}\cdot\mathbf{v})+r^2(\mathbf{r}\times\mathbf{\Omega}+\mathbf{v})],
\end{equation}
and
\begin{equation}\label{mignardT}
\mathbf{T}_{{\textrm{\scriptsize tide}}}=3k_2\Delta t\frac{GM^2R^5}{r^{8}}[-r^2\mathbf{\Omega}+\mathbf{r}\times\mathbf{v}],
\end{equation}
where $\mathbf{v}=\dot{\mathbf{r}}$ and $k_2$ is the second degree Love number. In this paper, we only consider the tides raised by the star on the planet, neglecting the tides raised on the star by the planet. Indeed, the strength of stellar tide is proportional to the planet mass and can be safely neglected in the case of super-Earths (see Rodr\'iguez \& Ferraz-Mello 2010). Note however that, after circularization of the planet orbit, stellar tides may be the only source of orbital decay.


In the reference frame centered in the star (see Figure \ref{elips}), the exact equations of motion of the planet can be now written as
\begin{eqnarray}\label{mov1}
&&\ddot{\mathbf{r}}=-\frac{G(M+m)}{r^3}\mathbf{r}
+\frac{(M+m)}{Mm}(\mathbf{F}_{{\textrm{\scriptsize tide}}}+\mathbf{F}_{22}),\nonumber\\
&&\ddot{\theta}=\dot{\Omega}=\frac{1}{C}(T_{{\textrm{\scriptsize tide}}}+T_{22}).
\end{eqnarray}
The perturbing forces $\mathbf{F}_{22}$ and $\mathbf{F}_{{\textrm{\scriptsize tide}}}$ alter the Keplerian orbit of the planet which is defined by the first term of the first equation, while the corresponding torques in the second equation produce variations of the planet rotation. Note that, in the planar case, this last equation can be written in scalar form, because both torques are normal to the orbital plane. 


\section{Numerical simulations}\label{num}

In this section we apply the analytical model described in the previous section, in order to investigate the motion of three known super-Earths planets, namely Kepler-10\,\textbf{b}, 55 Cnc\,\textbf{e} and GJ 3634\,\textbf{b}. For this task, the exact equations of motion (\ref{mov1}) were integrated numerically using the RA15 code (Everhart 1985). We stress that no scaling parameter, often used to accelerate the tidal evolution, was used in our simulations (see, for instance, Tittemore \& Wisdom (1988), for the validity of the usage of the scaling parameter).

Table 1 shows the physical parameters and current orbital elements of the systems considered in this paper.
\begin{tiny}
\begin{table}\label{tab1}
\begin{center}
\caption{\small Physical parameters and current orbital elements of the analysed systems ($^a$Batalha et al. 2011; $^b$Gillon et al. 2012; $^c$Bonfils et al. 2011). The first two planets were discovered through the transit technique, allowing us a good estimation of their radii. $m_{\oplus}$ and $R_{\oplus}$ stand for mass and radius of the Earth, respectively.}
\begin{tabular}{|c|c|c|c|c|c|}
\hline
   System & $M(\textrm{m}_{\odot})$ & $m(m_{\oplus})$ & $R(R_{\oplus})$ & $a_{{\textrm{\scriptsize current}}}$ (au)& $e_{{\textrm{\scriptsize current}}}$ \\
\hline
 Kepler-10$^a$ & $0.895$ & $4.56$ & 1.416 & 0.01684 & 0\\
 55 Cnc$^b$ & 0.905 & 8.63 & 2.17 & 0.0156 & 0.057\\

 GJ 3634$^c$ & 0.45 & 7.0 & - & 0.0287 & 0.08\\

  \hline
\end{tabular}
\end{center}
\end{table}
\end{tiny}

\subsection{Setting initial configurations}\label{sett}

Numerical integrations require a setting of the initial configurations of the systems. In the case of problems including dissipative processes, the choice is complicated, mainly, due to the inadequacy of the simulations of the past evolution of dissipative systems through integrations backward in time (see details in Michtchenko \& Rodr\'iguez 2011). However, as shown in that paper, the final state of a migrating system is independent of its initial position in the phase space, defined by three values of parameters: the mass and semi-major axes ratios and the total angular momentum. As a consequence, to assess the orbit configuration in the past, we have to set solely the starting semi-major axis value. The initial eccentricity should satisfy the conservation of the total angular momentum of the system.

The total angular momentum of the system is composed of the orbital component written as $m\sqrt{G(M+m)}\sqrt{a(1-e^2)}$, and the rotational component given by $C\Omega=\xi mR^2\Omega$. Using Kepler's third law, it is easy to show that the ratio of rotational and orbital components is of the order of $(R/a)^2$ (see Rodr\'iguez et al. 2011a); since $R\ll a$, the contribution of rotation to the total angular momentum can be neglected. Therefore, considering that the orbital angular momentum is nearly conserved during tidal migration, we obtain that
\begin{equation}\label{am}
e_{\textrm{\scriptsize ini}}\simeq\sqrt{1-\frac{a_{\textrm{\scriptsize current}}}{a_{\textrm{\scriptsize ini}}}(1-e_{\textrm{\scriptsize current}}^2)},
\end{equation}
where the subscripts ``ini" and ``current" stand for initial and current orbital elements, respectively. In this way, choosing an initial value for the semi-major axis and knowing the current values of $a$ and $e$, equation (\ref{am}) allows us to constrain the initial value of the eccentricity (see next section and specially Figure \ref{am-fig}). It should be emphasized that the value obtained in such a way must be considered as an approximation of the initial value of the eccentricity, since the planets may belong to systems with two or more planets. 

The value of $\Delta t$ is determined by the relationship between the tidal frequency $\nu$ and the quality factor $Q$. In the case of a linear approximation, we have $1/Q=\nu\Delta t$, where $\Delta t=$\,const. It is known that, in the Fourier spectrum of the tidal potential, the most important frequencies (that is, those involved in terms of order zero and order one in the eccentricity) are $\nu_0=2\Omega-2n$, $\nu_1=2\Omega-3n$ and $\nu_2=2\Omega-n$ . The first one is associated to the term which is independent of the eccentricity, while the two last frequencies correspond to terms which are proportional to $e$ (Darwin, 1879; Ferraz-Mello et al. 2008). Using the frequency $\nu_0$ and a modified version of the quality factor written in terms of the Love number $k_2$ as $Q'=3Q/2k_2$, we obtain
\begin{equation}\label{k2dt}
k_2\Delta t=\frac{3}{4Q'(\Omega-n)}.
\end{equation}

Actually, the spin rotation of exoplanets is not available from observations, thus we choose its initial value arbitrarily but considering a super-synchronous motion ($\Omega>n$).

Finally, according to Callegari \& Rodr\'iguez (2012), the equilibrium shape attained by the planet due to the gravitational interaction with the star, results in a prolateness, or equatorial ellipticity, which can be computed as $(B-A)/C\simeq(15/4)(M/m)(R/r)^3$. Knowing that $C_{22}=(\xi/4)(B-A)/C$ we can obtain the magnitude of $C_{22}$ for the planets with estimated radius. For the planets Kepler-10\,\textbf{b} and 55 Cnc\,\textbf{e} (see Section \ref{k10} and \ref{55cnc}), we obtain $C_{22}\simeq10^{-3}$ and $C_{22}\simeq2.4\times10^{-3}$, respectively, assuming $r=a_{{\textrm{\scriptsize current}}}$ . For sake of comparison with some rocky Solar System tidally evolved bodies, we note that $C_{22}=8.1\times10^{-6}$ in the case of Mercury (Smith et al. 2012 ), $C_{22}=1.0\times10^{-5}$ for Titan (Iess et al. 2010) and $C_{22}=5.6\times10^{-4}$ for Io (Schubert et al. 2004).

\subsection{Results}\label{k10-res}

The integration of the exact equations of motion usually requires a long computation time. For this reason, we decided to work with a small set of initial spin-orbit configurations and physical parameters. A statistical analysis taking into account many different initial conditions could give us a more accurate global picture of the past spin-orbit evolution of the planets. As pointed out by Correia \& Laskar (2009), such statistical study is more adequate in the context of the averaged equations of the spin-orbit problem (see Section \ref{mean}).

\begin{figure}
\begin{center}
\includegraphics[height=0.5\columnwidth,angle=270]{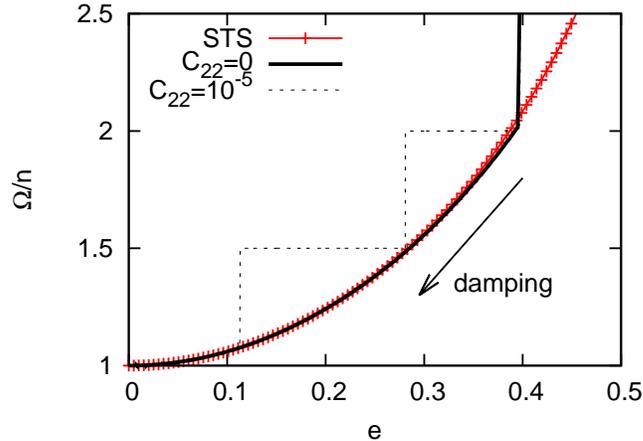}
\caption{\small The variation of the rotation with eccentricity of the Kepler-10\,\textbf{b} orbit. The arrow indicates the direction of the eccentricity and rotation evolutions. The curve with segments line shows the stationary tidal solution (STS) of the rotation, indicating a good agreement with the result of numerical simulation for $C_{22}=0$ (solid black curve). The dotted line corresponds to the case $C_{22}=10^{-5}$, where the trappings 2/1 and 3/2 occur. After escaping from the 3/2 resonant motion, the rotation follows the STS towards a synchronization between the rotation and orbital motions.}
\label{stat-k10}
\end{center}
\end{figure}

\subsubsection{Kepler-10\,\textbf{b}}\label{k10}

Kepler-10\,\textbf{b} is a very close-in planet with a period of $\sim$ 0.84 days, orbiting an old ($\simeq$ 12 Gyr) Sun-like star (Batalha et al. 2011). The physical parameters and the current orbital elements of the planet are summarized in Table 1. The estimated mass and radius values ($m=4.56m_{\oplus}$, $R=1.416R_{\oplus}$) suggest a rocky composition of the planet, which can be considered as a super-Earth (Valencia 2011). The transit technique measurements do not allow a reliable estimation of the current orbital eccentricity, which was arbitrarily fixed at zero in this work. This choice is supported by our numerical simulations of the whole system, which includes also the exterior companion, Kepler-10 \,\textbf{c}. The results have shown that the long-term evolution, due to the coupled effect of tides and mutual interaction, could result in the complete circularization of both planetary orbits (Rodr\'iguez et al., in preparation).

We adopt the initial semi-major axis at $a_{\textrm{\scriptsize ini}}=0.02$ au, which provides $e_{\textrm{\scriptsize ini}}=0.397$ through equation (\ref{am}). In addition, we set $\Omega_{\textrm{\scriptsize ini}}=2.7\,n_{\textrm{\scriptsize ini}}$ and $\xi=0.35$. For the tidal parameter, we assume $Q'=100$ (a value usually adopted for rocky planets) and, solving equation (\ref{k2dt}), we obtain $k_2\Delta t \simeq$ 1.1 min.

In Section \ref{sett} we have estimated that, for Kepler-10\,\textbf{b} on its current orbit, $C_{22}\simeq10^{-3}$. It should be emphasized that $C_{22}$ would be smaller if the frozen-in shape of the planet was attained at larger distances than $a_{{\textrm{\scriptsize current}}}$. Hence, in order to investigate different possible scenarios, we consider three different values of $C_{22}$, namely, $10^{-3}$, $10^{-4}$ and $10^{-5}$.
\begin{figure}
\begin{center}
\includegraphics[height=0.65\columnwidth,angle=270]{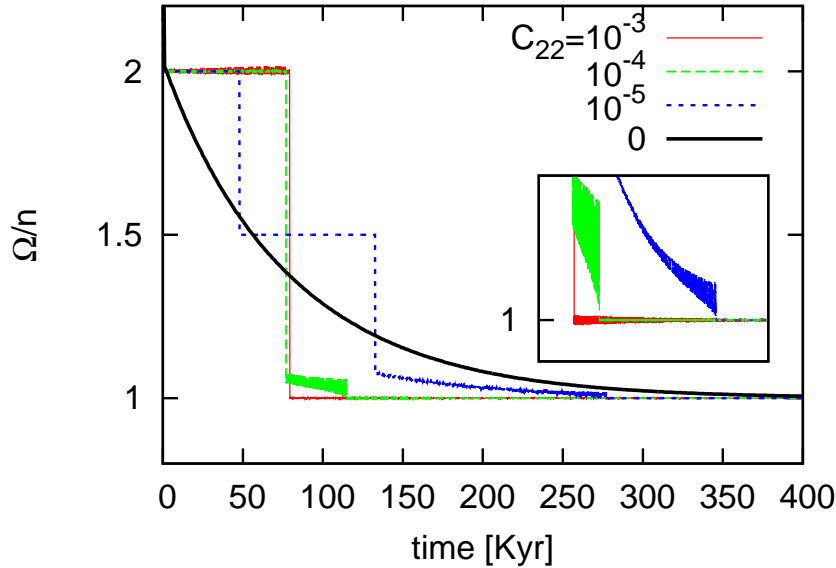}
\caption{\small Time variation of the ratio $\Omega/n$ for Kepler-10\,\textbf{b} depending on the value of $C_{22}$. Different evolutions are observed, including captures in 2/1, 3/2 and 1/1 spin-orbit resonances (SORs). The solid black curve corresponds to the evolution for $C_{22}=0$ (i.e., when only the tidal torque exist) and coincides with the stationary tidal solution (STS, see text for discussion) of the rotation. The captures in the synchronous motion are detailed on the small panel.}
\label{rot-k10}
\end{center}
\end{figure}

Figure \ref{stat-k10} shows the variation of $\Omega/n$ as a function of the eccentricity of Kepler-10\,\textbf{b}. The curve with segments line corresponds to the stationary tidal solution (hereafter STS) of the rotation ($\dot\Omega=0$), obtained when only the tidal torque is considered. The STS is a function which depends only on the planet eccentricity\footnote{It is worth noting that the STS does not depend on any physical parameter of the interacting bodies. However, alternative tidal theories show that the STS would depend on the internal viscosity (see Ferraz-Mello 2012).}, and is given by 

\begin{equation}\label{sync}
\Omega_{{\textrm{\scriptsize stat}}}/n=1+6e^2+3e^4/8+223e^6/8+{\cal O}(e^7)
\end{equation}
(see equations (42) or (43) of Hut (1981); see also Section \ref{mean} for more details). The stationary rotation evolves with the tidally damped eccentricity and the system reaches synchronization ($\Omega=n$) only for circular orbits. Note that for small values of $e$, there follows that $\Omega\sim n$; for this reason, the STS is also called pseudo-synchronous solution (but this is valid only for $e\ll1$).  

The solid black curve in Figure \ref{stat-k10} corresponds to the result of the numerical simulation performed with $C_{22}=0$ (only tidal forces are considered). In this case, the stationary rotation is attained almost instantaneously (when compared to the time of circularization). Note that both curves are in good agreement, at least for $e < 0.4$; hence, in this eccentricity interval, we can refer to the solution obtained with $C_{22}=0$ as STS. We also note that the curve for $C_{22}=0$ reaches the STS for $\Omega/n$ close to two; however, this is incidental and unrelated with the 2/1 trapping.


Figure. \ref{rot-k10} shows the time variation of $\Omega/n$. For the three considered values of $C_{22}>0$, the rotation rapidly encounters the value $\Omega/n=2$ and thus the planet rotation is captured in the 2/1 spin-orbit resonance (hereafter SOR). On one hand, for $C_{22}=10^{-3}$ and $C_{22}=10^{-4}$, the rotation escapes from the 2/1 trapping approximately at 80 Kyr, when the rotation drops to the value $\Omega/n=1$, implying capture into 1/1 SOR and synchronous motion. However, it is worth noting that, for $C_{22}=10^{-3}$, the synchronism is attained in a form of jump, whereas for $C_{22}=10^{-4}$ the rotation evolves without capture for a very short time before the 1/1 SOR (see small panel in Fig. \ref{rot-k10}). We have also plotted the result for $C_{22}=0$, showing that synchronization is attained around 400 Kyr. It is interesting to note that the introduction of the equatorial asymmetry of the planet reduces the time to reach synchronization; for instance, for $C_{22}=10^{-3}$, it is reduced by a factor of 5.	

On the other hand,  when $C_{22}=10^{-5}$, the 2/1 SOR is destabilized at around 48 Kyr giving rise to the trapping in the 3/2 SOR ($\Omega/n=3/2$). Thus, the permanence of the rotation in the 2/1 SOR is temporary and its duration clearly depends on the $C_{22}$ value. The rotation evolves in a Mercury-like state for some time, when a new instability occurs, after which an evolution free of trapping starts. Note that during the period of non-capture, the rotation returns to the stationary value, as can be seen in the dotted curve of Fig. \ref{stat-k10} when $e\leq0.11$. Finally, the synchronization is reached at around 270 Kyr.

\begin{figure}
\begin{center}
\includegraphics[height=0.85\columnwidth,angle=270]{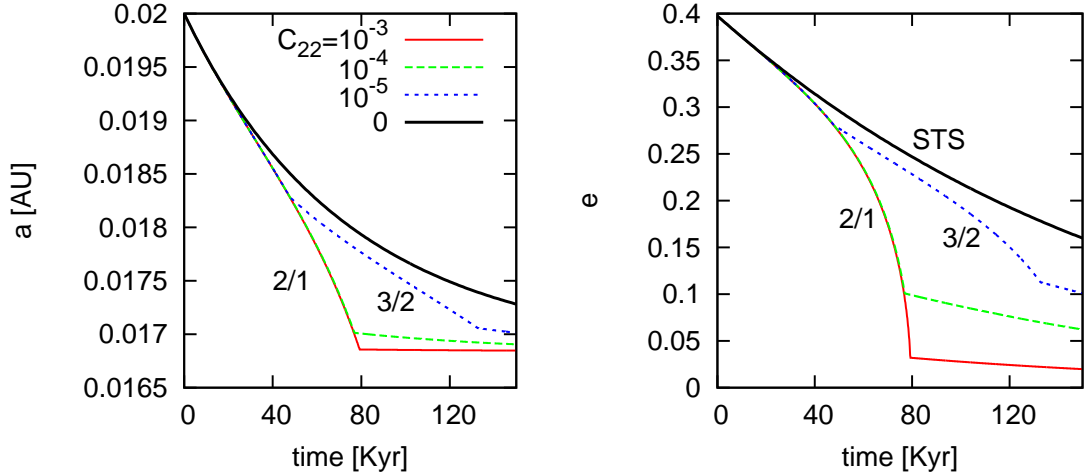}
\caption{\small Time variation of semi-major axis (left panel) and eccentricity (right panel) of Kepler-10\,\textbf{b} along the first 150 Kyr of evolution. The variation rates of $a$ and $e$ depend on the specific capture in which the rotation is trapped in. As different values of $C_{22}$ lead to several captures, different scenarios are expected for the orbital evolution, despite the final values of the elements being the same for all $C_{22}$ (see next figure).}
\label{ae-zoom-k10}
\end{center}
\end{figure}

\begin{figure}
\begin{center}
\includegraphics[height=0.85\columnwidth,angle=270]{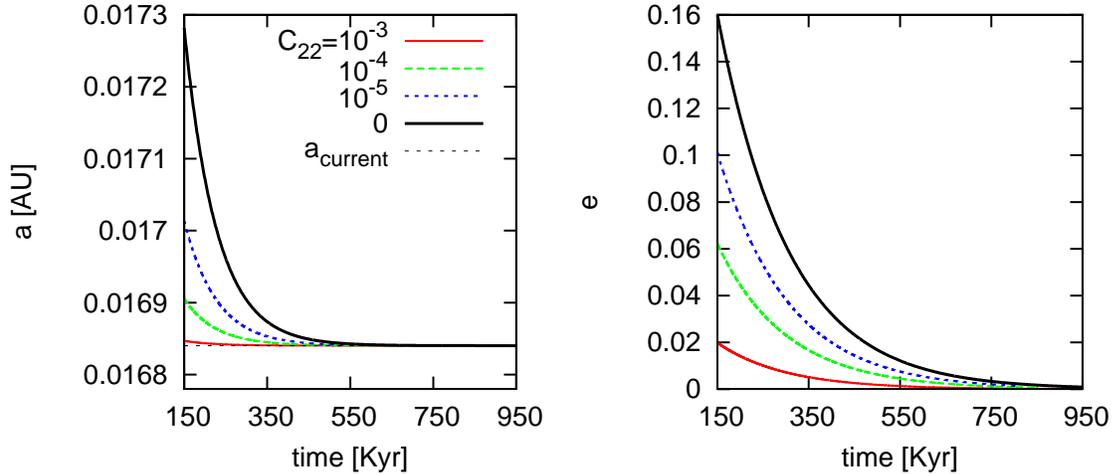}
\caption{\small Continuation of Figure \ref{ae-zoom-k10} shows the final evolution. The orbital decay continues until circularization is achieved. The current semi-major axis of Kepler-10\,\textbf{b} coincides with the one obtained in the simulation.}
\label{ae-k10}
\end{center}
\end{figure}

Figures \ref{ae-zoom-k10}--\ref{ae-k10} show the time evolution of semi-major axis and eccentricity of Kepler-10\,\textbf{b}. We note that, when the rotation evolves under capture in resonance (that is, for $C_{22}>0$), the rates of variations of the elements are larger than the one corresponding to the pure tidal case ($C_{22}=0$), with the difference increasing as the eccentricity is damped (when the comparison is done for a given SOR). Moreover, for the same value of $C_{22}$, the rate of orbital decay and circularization seems to depend on the specific SOR in which the planet rotation is trapped in. For example, in the case of $C_{22}=10^{-5}$, the evolutions of $a$ and $e$ show two different regimes as the rotation evolves first in the 2/1 and then in 3/2 trappings (see Section \ref{mean}).

We note that the time necessary to get an orbit with $e=0.05$ is much greater for $C_{22}<10^{-3}$. In fact, when the rotation evolves without capture, the time-scale for circularization increases, according to Figures \ref{ae-zoom-k10} and \ref{ae-k10}.

It is worth noting that, in the case of $C_{22}>0$, the synchronous motion can be attained even for $e\neq0$. For instance, in the case of $C_{22}=10^{-4}$, the system is captured in 1/1 SOR at roughly 120 Kyr, when $e\simeq0.079$. This fact confirms the classical result that, for eccentric orbits, the synchronous motion is only possible if there exists an additional torque which counteracts the tidal torque (see Goldreich \& Peale 1966; Ferraz-Mello et al. 2008 and references therein). We also note that after circularization, the 1/1 SOR is the only possible resonant motion\footnote{Except if an additional mechanism, such as impact events, acts to destabilize the synchronous motion (see Wieczorek et al. 2012; Correia \& Laskar 2012)}.

\begin{figure}
\begin{center}
\includegraphics[height=0.5\columnwidth,angle=270]{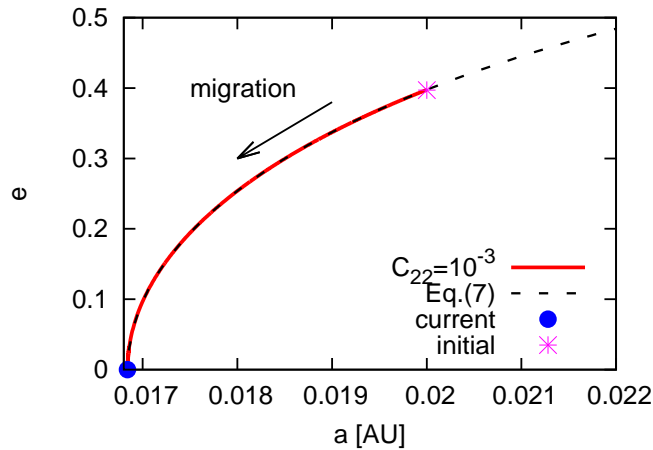}
\caption{\small Evolution of the orbit of Kepler-10\,\textbf{b} in the plane $(a,e)$. The solution of equation (\ref{am}) is shown by the dashed line and is compared to the result of the numerical simulation for the case $C_{22}=10^{-3}$, giving a good agreement (the curves for the other two $C_{22}$ values also agree with the solution of equation (\ref{am}), but are not shown here for sake of simplicity). The starting position of the planet is indicated by the star symbol.}
\label{am-fig}
\end{center}
\end{figure}

Figure \ref{am-fig} shows the evolution of the system in the plane of orbital elements $a$ and $e$. The solid line shows the result of numerical simulation in the case of $C_{22}=10^{-3}$. The dashed line corresponds to equation (\ref{am}), which is independent on the value of $C_{22}$. Note as both curves are in very good agreement, in such a way confirming our assumption about the conservation of the orbital angular momentum (Section \ref{sett}).

\subsubsection{GJ 3634\,\textbf{b}}\label{gj3634}

GJ 3634 b is a super-Earth planet with mass $m=7.0m_{\oplus}$ orbiting a M2.5 dwarf each 2.65 days with an eccentricity of $e=0.08$ (Bonfils et al. 2011). Since there is no transit detection for this planet, its radius is unknown. However, Valencia, Sasselov \& O'Connell (2007) have shown that there is a maximum radius that a rocky terrestrial planet may achieve for a given mass. A larger radius would imply a large amount of water and even the possibility of a massive volatile envelope. Indeed, for a planet with mass $7.5m_{\oplus}$ the maximum terrestrial radius would be 11,600 km ($\simeq1.82R_{\oplus}$), which result in a mean density of $\rho\simeq1.25\rho_{\oplus}$. Hence, we assume for GJ 3634 b that $\rho=\rho_{\oplus}$ to obtain $R\simeq1.9R_{\oplus}$. 

The adopted initial orbital elements are $a_{\textrm{\scriptsize ini}}=0.033$ au and $e_{\textrm{\scriptsize ini}}=0.369$, while $\Omega_{\textrm{\scriptsize ini}}=3.7\,n_{\textrm{\scriptsize ini}}$, $\xi=0.35$ and $Q'=100$ ($k_2\Delta t=2.1$ min). At the current value of the semi-major axis the estimated value of $C_{22}$ is $C_{22}\simeq1.6\times10^{-4}$ (Section \ref{sett}); in this work, we perform the analysis using two values of $C_{22}$: $10^{-4}$ and $10^{-5}$.

\begin{figure}
\begin{center}
\includegraphics[height=0.475\columnwidth,angle=270]{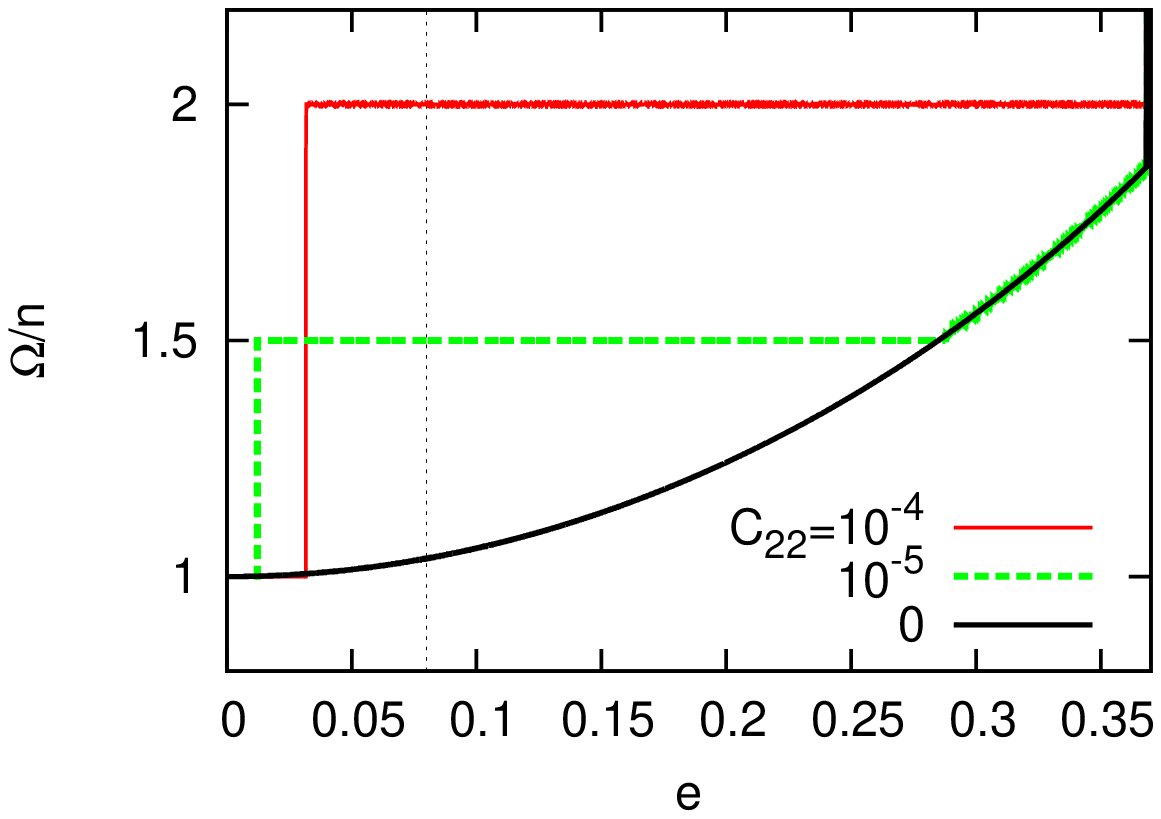}
\includegraphics[height=0.475\columnwidth,angle=270]{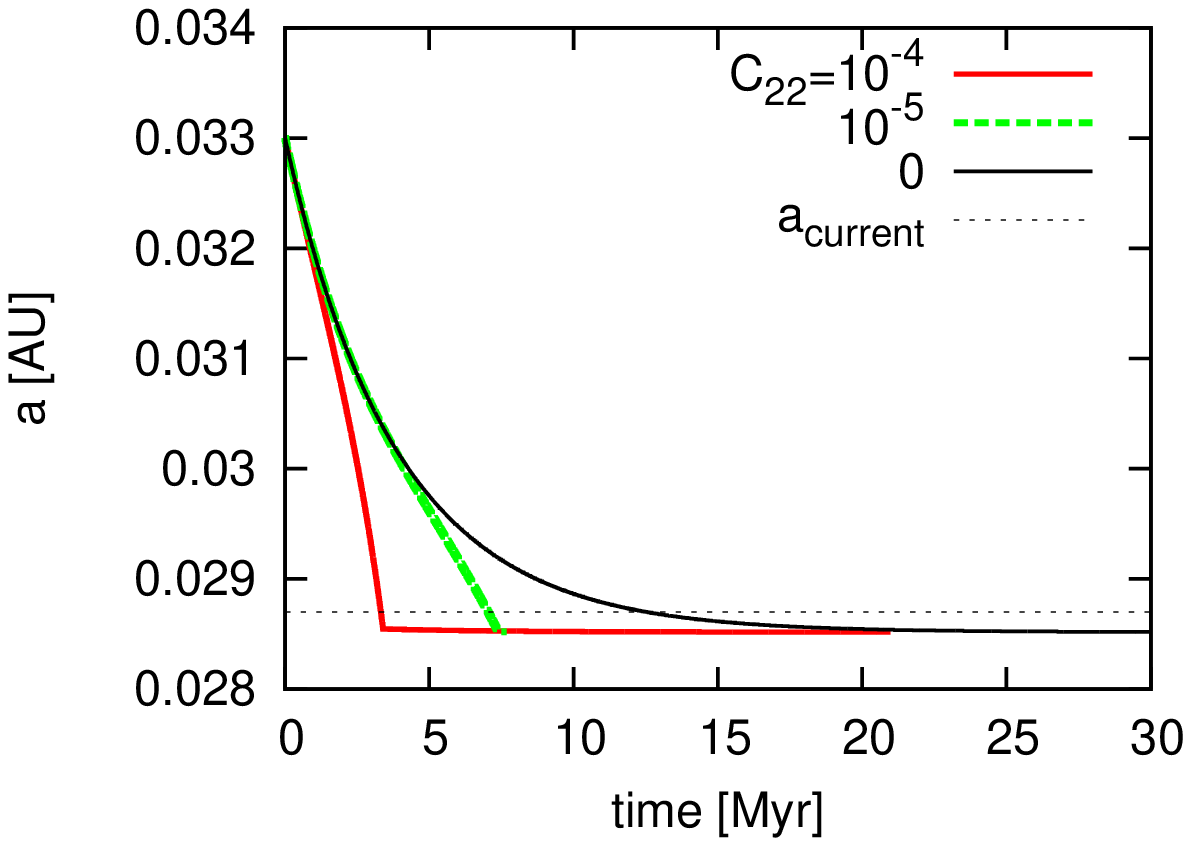}
\caption{\small The variation of $\Omega/n$ as a function of $e$ (left panel) and the time evolution of $a$ (right panel) for the planet GJ 3634\,\textbf{b}. Depending on $C_{22}$, the planet rotation may be currently evolving in either 2/1 or 3/2 SOR. Note that the time-scale of orbital decay is drastically reduced provided the planet can support a permanent deformation resulting in a nonzero $C_{22}$ value. The vertical and horizontal dotted lines indicates the current values of $e$ and $a$, respectively}
\label{rot-exce-gj3634}
\end{center}
\end{figure}

Figure \ref{rot-exce-gj3634} shows the variation of $\Omega/n$ as a function of $e$ (left panel) and the time variation of $a$ (right panel). For $C_{22}=10^{-4}$, the rotation is rapidly trapped in the 2/1 SOR, remaining in such resonant configuration during the most time of the tidal eccentricity damping. Only at low eccentricities, close to 0.032, the system leaves the 2/1 SOR dropping directly into the synchronous motion.

At variance, for $C_{22}=10^{-5}$, the rotation quickly reaches the stationary (non-resonant) state given by the STS curve ($C_{22}=0$) in Figure \ref{rot-exce-gj3634} (left). The system evolves along the STS curve up to close approach to the 3/2 SOR. Then, the system is captured and remains in this resonance until dropping in the synchronized state, at $e\simeq0.012$.

In both cases, the final configuration of the planet's orbit is characterized by circularization and synchronization, as in the case of Kepler-10\,\textbf{b}. However, the current eccentricity of the GJ 3634\,\textbf{b} is different from zero (its location at 0.08 is shown by the dotted vertical line in the left panel of Figure \ref{rot-exce-gj3634}). Note that the current eccentricity is reached when the rotation still evolves in the 2/1 or 3/2 SOR (depending on $C_{22}$). This fact suggests that GJ 3634\,\textbf{b} could be presently involved in one of such resonant trappings, in particular, it could be found in a Mercury-like rotational state ($\Omega/n\simeq3/2$). However, it should be kept in mind, that this conclusion depends on the choice of specific values of $C_{22}$ and $Q'$. Indeed, the results of an additional simulation performed with $C_{22}=10^{-6}$, have shown that the trapping of the system in the 3/2 SOR is destabilized at $e\simeq0.11$. Since $e_{{\textrm{\scriptsize current}}}=0.08<0.11$, the planet rotation would be currently evolving without capture (along the STS) for such a small $C_{22}$ value. 

The semi-major axis evolution (Figure \ref{rot-exce-gj3634}, right panel) shows that the tidal decay is faster when the rotation of the planet is trapped in the 2/1 and 3/2 spin-orbit resonant motions. The fact that the current eccentricity of the planet orbit is different from zero suggests that, in the absence of other perturbations, the orbital decay\footnote{In fact, $\dot{a}\propto e^2$ when $\Omega=n$ or $\Omega\simeq n$ (see equation (\ref{a-11-tot2})).} of the planet is presently in progress and, consequently, the final semi-major axis will be smaller than the current one.

\subsubsection{55 Cnc\,\textbf{e}}\label{55cnc}

55 Cnc\,\textbf{e} is a hot super-Earth planet, with the mass $m=8.63m_{\oplus}$ and the radius $R=2.17R_{\oplus}$, orbiting a Sun-like star each 0.74 days with $e=0.057$ (Winn et al. 2011; Gillon et al. 2012). The averaged density of the planet is comparable to that of Earth, suggesting a rock-iron composition supplemented by a significant mass of water, gas, or other light elements\footnote{The physical composition and the suggested value of $Q'$ ($5.5\times10^3$) indicate that the possibility of 55 Cnc\,\textbf{e} to be a mini-Neptune instead of a super-Earth should not be ruled out.} (Winn et al. 2011).

For numerical simulations, we adopt $a_{\textrm{\scriptsize ini}}=0.0186$ au, $e_{\textrm{\scriptsize ini}}=0.4$, $\Omega_{\textrm{\scriptsize ini}}=2.7\,n_{\textrm{\scriptsize ini}}$ and $\xi=0.35$. The tidal history of the system indicates that the magnitude of $Q'$ must be higher than the commonly adopted values for rocky planets (Ferraz-Mello 2012). In this work, we use $Q'=5.5\times10^3$; this value allows us to obtain $k_2\Delta t\simeq1.1$ s. We perform numerical simulations for two different values of $C_{22}$: $10^{-3}$ and  $10^{-4}$. For the purpose of further comparison with the results reported in Callegari \& Rodr\'iguez (2012), we also consider the values $C_{22}=2.45\times10^{-3}$ and $C_{22}=2.45\times10^{-4}$.

\begin{figure}
\begin{center}
\includegraphics[height=0.6\columnwidth,angle=270]{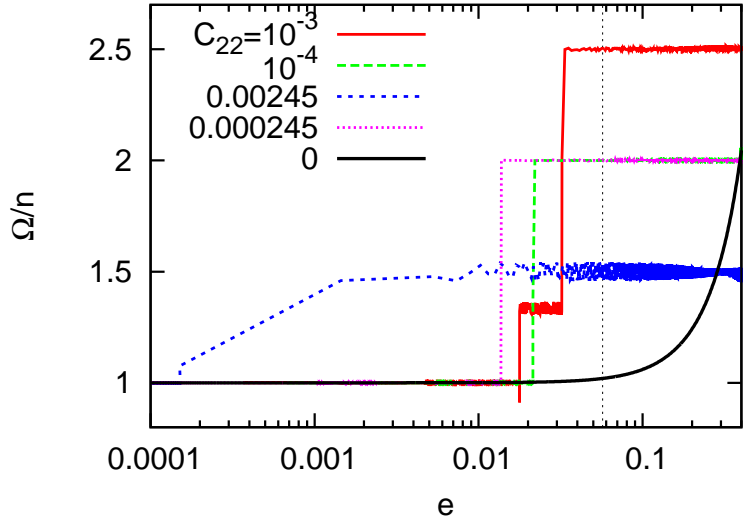}
\caption{\small The variation $\Omega/n$ as a function of $e$ (in logarithmic scale), for the planet 55 Cnc\,\textbf{e}. We have used $C_{22}=10^{-3}$ in addition to the value adopted in Callegari \& Rodr\'iguez (2012) $C_{22}=2.45\times10^{-3}$. Also values ten times smaller were taken into account for sake of completeness. The vertical dotted line indicates the current value of $e$, namely, 0.057. The trappings in the RSOs  5/2, 2/1, 3/2, 4/3 and 1/1 appear during the evolutions. Note the large amplitude oscillation of $\Omega/n$ inside the 3/2 and 4/3 cases with $C_{22}$ of the order of $10^{-3}$. Depending on $C_{22}$, 55 Cnc\,\textbf{e} can be an example of a planet currently evolving in the 5/2, 2/1 or 3/2 SOR.}
\label{rot-exce-55cnc}
\end{center}
\end{figure}

Figure \ref{rot-exce-55cnc} shows the evolution of rotation as a function of the planet's eccentricity. The captures inside the 5/2, 2/1, 3/2, 4/3 and 1/1 resonances can be observed. For $C_{22}=2.45\times10^{-3}$, the planet rotation evolves under capture in the 3/2 SOR almost up to the orbital circularization, with $\Omega/n$ roughly varying between 1.46 and 1.54.

\begin{figure}
\begin{center}
\includegraphics[height=0.6\columnwidth,angle=270]{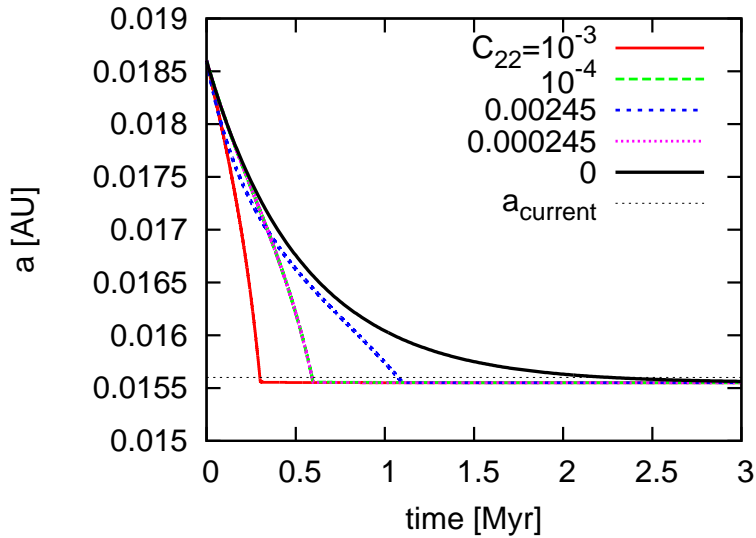}
\caption{\small Time variation of semi-major axis for 55 Cnc\,\textbf{e}. The curves show the evolution for four values of $C_{22}$ together with the STS curve. As for previously analysed planets, the orbital decay is faster when the planet rotation is trapped in a resonant motion. The current value of $a$ is obtained for all values of $C_{22}$.}
\label{elem-55cnc}
\end{center}
\end{figure}

In a recent work, Callegari \& Rodr\'iguez (2012) have shown that, for $e=0.057$ and $(B-A)/C=0.028$ (or $C_{22}=2.45\times10^{-3}$), the phase space of the rotation ($\dot{\theta}/n,\theta$) of 55 Cnc\,\textbf{e}, between the main spin-orbit resonances, is dominated by a large region of chaotic motion. This region increases as larger values of the eccentricity are considered.

Capture in spin-orbit resonance can work as a mechanism to prevent a chaotic rotation, provided that the dissipation is large enough in order to ensure the corresponding trapping. According to our simulations (see Figure \ref{rot-exce-55cnc}), destabilizations of different resonances occur as the eccentricity is tidally damped (see also next section), giving rise to new temporary trappings with a subsequent evolution to synchronism. In this way, a chaotic behavior of the rotation can be avoided by a continuous evolution under resonant motion. It is important to note that, at least for the cases shown in Figure \ref{rot-exce-55cnc}, there is no interval of non-resonant motion between two successive trappings, as in the case of Kepler-10\,\textbf{b} (see Figure \ref{rot-k10} for $C_{22}=10^{-4}$ and $10^{-5}$).




An additional simulation done with $C_{22}=2.45\times10^{-5}$ (not shown in figures), has resulted in a 2/1 trapping until $e\simeq0.04$, indicating that the planet rotation may be currently evolving in resonant motion, for such a small value of $C_{22}$.


Figure \ref{elem-55cnc} shows the time variation of semi-major axis. It is worth noting that the orbital decay results in a more rapid evolution towards $a_{{\textrm{\scriptsize current}}}$ when the planet rotation is trapped in a SOR. Also, the rate of variation depends on the specific resonance. From Figure \ref{elem-55cnc} we see that the evolution when $C_{22}\sim10^{-4}$, (capture in 2/1) is almost two times faster than the one for $C_{22}=2.45\times10^{-3}$ (3/2 capture). We will return to this discussion in the next section.




\section{Averaged equations}\label{mean}

In order to better understand the results previously obtained through numerical simulations of the exact equations of motion, in this section, we develop an analytical approach in the context of the averaged equations of the spin-orbit problem.

The equation of motion for the spin in equation (\ref{mov1}) cannot be solved analytically. However, because we are interested in the evolution inside spin-orbit resonances, we will consider planetary spin rates having values such that $\Omega\simeq pn$, where $p$ is a half-integer. Our investigation is restricted to the cases $p=1$, $p=3/2$ and $p=2$ but can be easily extended to any other SOR.

We introduce the ``resonant'' angle $\delta=\theta-p\,l$, where $l$ is the mean anomaly of the planet orbit and $\theta$ was defined in Section \ref{model}. Note that the angle $\delta$ gives the orientation of the long axis whenever the planet passes through the pericenter of its orbit (i.e., $\delta=\theta$ when $l=0$, see Figure \ref{elips}). It is easy to show that, inside the spin-orbit resonance given by the parameter $p$, the resonant angle is a slowly varying variable when compared to the orbital period of the planet. Indeed, we can write that $\dot{\theta}=\dot{\delta}+pn$, where $\dot{\theta}=\Omega\simeq pn$ and, consequently, $\dot{\delta}\simeq 0$. Now, replacing $\theta$ by $\delta+pl$ in equation (\ref{t22}) and averaging over the orbital period, we obtain the classical result for the averaged torque on the planet produced by the permanent equatorial deformation (see Goldreich 1966; Goldreich \& Peale 1966):
\begin{equation}\label{t22mean}
<T_{22}>=-6mR^2C_{22}n^2H(p,e)\sin2\delta,
\end{equation}
where $H(p,e)$ are power series of $e$ with the leading term of order $e^{2|p-1|}$ (see Appendix A for a literal expansion up to seventh order).  

The averaged tidal torque on the planet is given by
\begin{equation}\label{ttide-mean}
<T_{{\textrm{\scriptsize tide}}}>=K_{{\textrm{\scriptsize tide}}}\left(\frac{n}{a}\right)^3{\cal T}(p,e),
\end{equation}
where $K_{{\textrm{\scriptsize tide}}}=3k_2\Delta tMR^5$ and ${\cal T}(p,e)$ is a suitable function, whose expansion in powers of $e$ is shown in Appendix A (Hut 1981). 

At resonance, the equilibrium solution of the rotation is given by the condition 

\begin{equation}
C\dot{\Omega}=<T_{{\textrm{\scriptsize tide}}}>+<T_{22}>=0,\nonumber
\end{equation}
and, resolving for $C_{22}\sin2\delta$, we obtain
\begin{equation}\label{j22-gral}
C_{22}\sin2\delta=\frac{1}{2}k_2\Delta t\,n\frac{M}{m}\left(\frac{R}{a}\right)^3\frac{{\cal T}(p,e)}{H(p,e)}.
\end{equation}
If we know the maximum value that $\delta$ can reach without disturbing the trapping in the $p$--SOR, we may obtain the minimum value of $C_{22}$ necessary for capture in the same resonant motion. 


\subsection{1/1 Spin-Orbit Resonance}\label{1/1}

Replacing $p=1$ in equation (\ref{j22-gral}) and expanding the ratio $\frac{{\cal T}(p,e)}{H(p,e)}$ in powers of $e$, up to order 6, we obtain

\begin{equation}\label{j22-11}
C_{22}\sin2\delta=\frac{1}{16}k_2\Delta t\,n\frac{M}{m}\left(\frac{R}{a}\right)^3(48e^2+483e^4+2624e^6)+{\cal O}(e^8).
\end{equation}
We note that, for ${\cal O}(e^4)$, the above expression is the same obtained in Ferraz-Mello et al. (2008), for synchronous asymmetric planets \footnote{In this case, the association $\epsilon_2=(2\Omega-n)\Delta t=n\Delta t$ should be done, where $\epsilon_2$ is the phase lag of the tidal wave whose frequency is $(2\Omega-n)$.}.

\subsubsection{Semi-major axis evolution}\label{a-1/1}

The averaged variation of the orbital elements can be obtained using the Euler-Gauss equations (see Beutler 2005, p. 230), which are written in terms of the radial and transverse components of the perturbing force. The result for the semi-major axis is
\begin{equation}\label{a-11}
<\dot{a}_{22}>=-\frac{3nR^2C_{22}\sin2\delta}{72a}\left(-288+720 e^2-234 e^4+35e^6\right)+{\cal O}(e^8).
\end{equation}
Replacing $C_{22}\sin2\delta$ by the corresponding expression given in equation (\ref{j22-11}), we obtain
\begin{equation}\label{a-11-stat}
<\dot{a}_{22}>=3k_2\Delta t\frac{M}{m}\left(\frac{R}{a}\right)^5n^2a\left(12e^2+\frac{363}{4}e^4+\frac{2911}{8}e^6\right)+{\cal O}(e^8).
\end{equation}

In the context of the linear tidal model, the averaged variation of the semi-major axis due to tides is given by equation (9) of Hut (1981), valid for any values of $p$ and $e$. The expansion up to ${\cal O}(e^8)$ reads

\begin{equation}\label{a-tide-gral}
<\dot{a}_{{\textrm{\scriptsize tide}}}>=3k_2\Delta t\frac{M}{m}\left(\frac{R}{a}\right)^5n^2a\Big{[}2(-1+p)+(-46+27p)e^2+\frac{3}{4}(-480+191p)e^4+\frac{1}{8}(-13530+3961p)e^6\Big{]}+{\cal O}(e^8).
\end{equation}
Replacing $p=1$ in the above equation we obtain

\begin{equation}\label{a-11-tide}
<\dot{a}_{{\textrm{\scriptsize tide}}}>_{p=1}=-3k_2\Delta t\frac{M}{m}\left(\frac{R}{a}\right)^5n^2a\left(19e^2+\frac{867}{4}e^4+\frac{9569}{8}e^6\right)+{\cal O}(e^8).
\end{equation}

The total averaged variation of the semi-major axis is then given as

\begin{equation}\label{a-11-tot}
<\dot{a}>_{p=1}\,=\,<\dot{a}_{{\textrm{\scriptsize tide}}}>_{p=1}+<\dot{a}_{22}>
\end{equation}
and thus
\begin{equation}\label{a-11-tot2}
<\dot{a}>_{p=1}=-3k_2\Delta t\frac{M}{m}\left(\frac{R}{a}\right)^5n^2a\left(7e^2+\frac{504}{4}e^4+\frac{3329}{4}e^6\right)+{\cal O}(e^8).
\end{equation}
At the same order of approximation, the above equation coincides with the result of previous works (see Ferraz-Mello et al. (2008) and references therein).

When only the tidal torque on the planet is taken into account, the non-resonant STS is given by equation (\ref{sync}).
In this case, the synchronism is possible only for circular orbits. If instead of $p=1$, we replace $p=p_{{\textrm{\scriptsize stat}}}=\Omega_{{\textrm{\scriptsize stat}}}/n$ in equation (\ref{a-tide-gral}), we obtain


\begin{equation}\label{a-11-tide-stat}
<\dot{a}_{{\textrm{\scriptsize tide}}}>_{p=p_{{\textrm{\scriptsize stat}}}}=-3k_2\Delta t\frac{M}{m}\left(\frac{R}{a}\right)^5n^2a\left(7e^2+54e^4+\frac{1133}{4}e^6\right)+{\cal O}(e^8).
\end{equation}

We define $r_a^p$ as the ratio between averaged semi-major axis variation in the $p$--SOR ($p=1,3/2,2,...$) and in the tidal stationary solution of the rotation ($p=p_{{\textrm{\scriptsize stat}}}$), that is

\begin{equation}\label{ra}
r_a^p\equiv\frac{<\dot{a}>_{p}}{<\dot{a}_{{\textrm{\scriptsize tide}}}>_{p=p_{{\textrm{\scriptsize stat}}}}}.
\end{equation}
Comparing equations (\ref{a-11-tot2}) and (\ref{a-11-tide-stat}), we note that $r_a^{p=1}>1$, indicating that the rate of semi-major axis decay is stronger when the rotation is locked in the 1/1 SOR, compared to the case when it evolves following the STS. This result is in agreement with Wisdom (2008), who has shown that the rate of energy dissipation in a body with synchronous rotation is larger than in a asymptotic non-synchronous rotation.

\subsubsection{Eccentricity}\label{e-1/1}

We repeat the procedure described in the previous section to obtain the averaged variation of the planet eccentricity. The resulting expressions are:

 \begin{equation}
<\dot{e}_{22}>=-9k_2\Delta t\frac{M}{m}\left(\frac{R}{a}\right)^5n^2\left(e^3+\frac{117}{16}e^5\right)+{\cal O}(e^7),
\end{equation}

\begin{equation}
<\dot{e}_{{\textrm{\scriptsize tide}}}>_{p=1}=-\frac{3}{2}k_2\Delta t\frac{M}{m}\left(\frac{R}{a}\right)^5n^2\left(7e+113e^3+\frac{5299}{8}e^5\right)+{\cal O}(e^7).
\end{equation}

\begin{equation}
<\dot{e}_{{\textrm{\scriptsize tide}}}>_{p=p_{{\textrm{\scriptsize stat}}}}=-\frac{3}{8}k_2\Delta t\frac{M}{m}\left(\frac{R}{a}\right)^5n^2(28e+188e^3+917e^5)+{\cal O}(e^7).
\end{equation}

\begin{equation}
<\dot{e}>_{p=1}=-\frac{3}{8}k_2\Delta t\frac{M}{m}\left(\frac{R}{a}\right)^5n^2(28e+476e^3+2825e^5)+{\cal O}(e^7),
\end{equation}

Similarly to the semi-major axis variation case, $r_e^{p=1}>1$, where $r_e^{p}$ is the analogous for the eccentricity of equation (\ref{ra}).

\subsection{3/2 and 2/1 Spin-Orbit Resonances}\label{3/2-2/1}

We analyse in this section the spin-orbit behavior when the planet evolves in the 3/2 and 2/1 SORs, employing the same method described in the previous section\footnote{We stress that $H(p>1,e)$ has no term independent of $e$ and thus equation (\ref{j22-gral}) cannot be expanded around $e\simeq0$.}. We have



\begin{equation}\label{a-32-tot}
{<\dot{a}>_{p=3/2}}=-3k_2\Delta t\frac{M}{m}\left(\frac{R}{a}\right)^5n^2a\left(\frac{1}{2}-\frac{5}{4}e^2+\frac{774}{16}e^4+\frac{8019}{16}e^6\right)+{\cal O}(e^8),
\end{equation}
and

\begin{equation}\label{a-21-tot}
{<\dot{a}>_{p=2}}=-3k_2\Delta t\frac{M}{m}\left(\frac{R}{a}\right)^5n^2a\left(2-2e^2-3e^4+\frac{943}{4}e^6\right)+{\cal O}(e^8).
\end{equation}

\begin{figure}
\begin{center}
\includegraphics[height=0.35\columnwidth,angle=0]{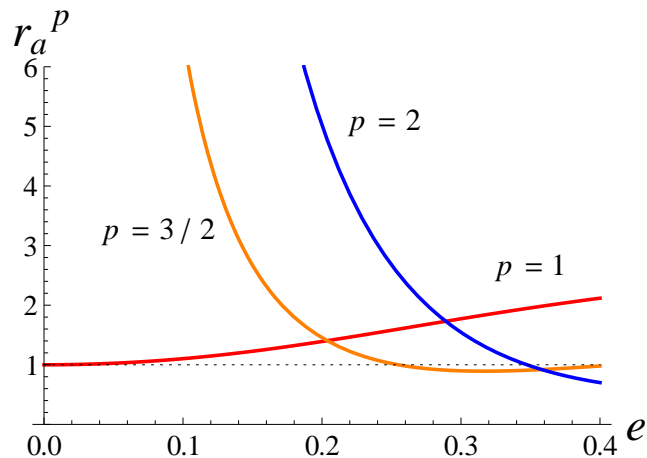}
\caption{\small The variation of $r_a^p$ as a function of orbital eccentricity, for $p=2$, 3/2 and 1. In the cases of $p>1$ the ratio of rates of variation of $a$ in the SOR and in the STS increases for small $e$, showing that the orbital decay is enhanced (when compared with the evolution in the STS) whenever the planet rotation evolves under resonance with the orbital motion.}
\label{razao}
\end{center}
\end{figure}

Figure \ref{razao} shows a plot of $r_a^p$ for the three analysed resonances as a function of the eccentricity. On one hand, we note that $r_a^{p=3/2}$ and $r_a^{p=2}$ increase for small eccentricity, indicating that the major effect on the orbital evolution due to the resonance captures occurs as $e$ decreases, in agreement with the results of the numerical simulations (see Figures \ref{ae-zoom-k10}, \ref{rot-exce-gj3634} and \ref{elem-55cnc}). On the other hand, for large $e$, $r_a$ is small (mainly in the case $p=3/2$) as also shown in Figures \ref{ae-zoom-k10} and \ref{rot-exce-gj3634}, where there is a small difference between the evolutions with and without $C_{22}$ (i.e, $r_a\simeq1$).

When the synchronous motion is analysed, we note in Figure \ref{razao} that $r_a^{p=1}$ is an increasing function of $e$, indicating that the orbital decay when the rotation is trapped in the 1/1 SOR is large for large $e$. However, as simulations show, synchronism is attained for very low values of $e$, making the orbital decay almost negligible.

The results above are also valid for $r_e^p$. It is interesting to note that $r_a^p$ and $r_e^p$ are only functions of the eccentricity and do not depend on any physical parameter.

It is worthwhile mentioning that $T_{22}$ averages to zero unless the condition $\Omega\simeq pn$ is assumed (whit $p$ integer), that is, the averaged orbital evolution due to the triaxial torque is expected only for resonant cases.  

\subsection{Critical $C_{22}$}\label{c22c}

As it has been noted, equation (\ref{j22-gral}) can be used to obtain the minimum value of $C_{22}$ necessary for the stability of the $p$--SOR. Calling $C_{22}^*$ the one value given by the right hand of equation (\ref{j22-gral}), the stability condition for a resonant spin state requires that $C_{22}>C_{22}^*$. The above condition is frequently found in literature as strength criterion of the resonance (see Murray \& Dermott 1999, p. 205).



\begin{figure}
\begin{center}
\includegraphics[height=0.4\columnwidth,angle=0]{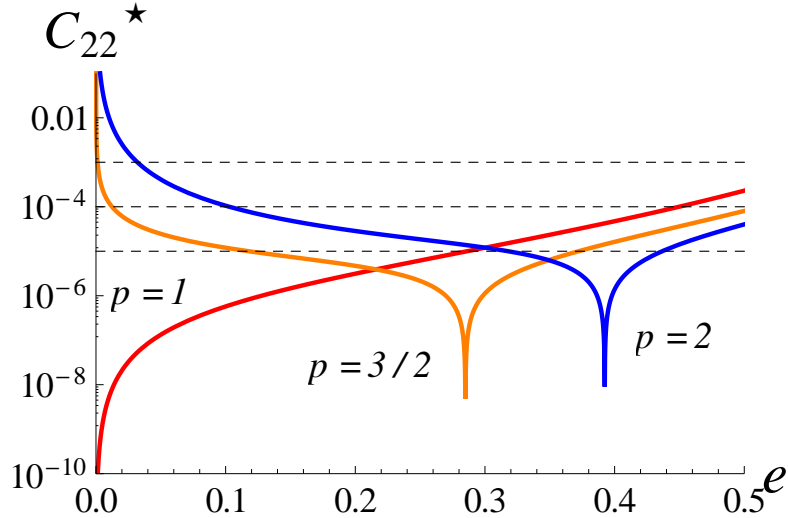}
\caption{\small Critical value of $C_{22}$ given by equation (\ref{j22-gral}) and calculated with parameters of the planet Kepler
-10\,\textbf{b}. The calculation of $C_{22}^*$ can be used as a guide to investigate the stability of a given SOR as a function of eccentricity (see text for discussion). Note that in the case of $p=1$ we have $C_{22}^*=0$ for $e=0$ (see equation (\ref{j22-11})), indicating that the synchronism is the only possible resonant motion for circular orbits (see Goldreich \& Peale 1966; Celletti \& Chierchia 2008; Callegari \& Rodr\'iguez 2012).}
\label{c22}
\end{center}
\end{figure}

Figure \ref{c22} shows the functional dependence of the absolute value of $C_{22}^*$ with the eccentricity, obtained for Kepler-10\,\textbf{b}. The dashed horizontal lines represent three values of $C_{22}$ used in the numerical simulations shown in Section \ref{k10}, namely, $10^{-3},10^{-4}$ and $10^{-5}$. 

\begin{figure}
\begin{center}
\includegraphics[height=0.6\columnwidth,angle=270]{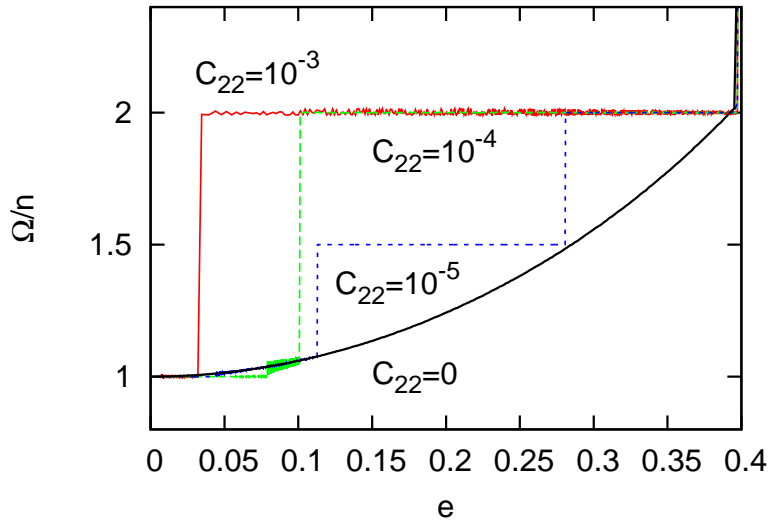}
\caption{\small The variation of $\Omega/n$ as a function of the orbital eccentricity for Kepler-10\,\textbf{b}, resulting from the analysis of numerical simulations of the exact equations of motion (see Section \ref{k10-res}). If a comparison with the results of Figure \ref{c22} is done, we  obtain a good agreement for the values of $e$ for which each SOR destabilizes (see text for details).}
\label{rot-exce-k10}
\end{center}
\end{figure}

The case $p=2$, shown by the blue curve, indicates that $C_{22}^*=10^{-5}$ for $e\simeq0.3$, $C_{22}^*=10^{-4}$ for $e\simeq0.1$ and $C_{22}^*=10^{-3}$ for $e\simeq0.03$. A good agreement is obtained doing the comparison with the results of the numerical simulations for Kepler-10\,\textbf{b}, shown in Figure \ref{rot-exce-k10}. Note as the 2/1 SOR destabilizes when $e\simeq0.28$, $e\simeq0.1$ and $e\simeq0.03$. 

The same can be observed for the 3/2 SOR, in which the rotation escapes from the trapping when $e\simeq0.11$ if $C_{22}=10^{-5}$, in concordance\footnote{Small differences between the result of numerical simulations and the analysis of averaged equations may arise because for the construction of Figure \ref{c22} we use $a=a_{{\textrm{\scriptsize current}}}$, taking also into account that truncated series in $e$ were used for the same figure.} with the result of Figure \ref{c22} (orange curve).

Since for roughly $e\leq0.1$ the condition $C_{22}>C_{22}^*$ is satisfied more easily for $p=1$ as compared to $p>1$, there is no problem in understanding the occurrence of synchronous motion for very small eccentricity, as can be also verified in Figure \ref{rot-exce-k10}.


It is worth noting that instead to compute a critical $C_{22}$ value, the condition for stationary solutions $<T_{{\textrm{\scriptsize tide}}}>+\\<T_{22}>=C\dot{\Omega}=0$ also enables us to obtain a critical value of $e$ for a given fixed $C_{22}$. The resonant motion would destabilize whenever the eccentricity becomes smaller than the critical value for each SOR. This approach is more convenient in those cases for which the observations allow an estimation of $C_{22}$ (see Correia \& Laskar (2004) for the study of the rotation of Mercury).

\section{Discussion and conclusions}\label{discussion}

The numerical exploration of the planet rotation which is subject to the conservative triaxial gravitational torque of the star shows different behaviors, including the oscillation around the equilibrium points of spin-orbit resonances (see Callegari \& Rodr\'iguez 2012 and references therein). When dissipative effects produced by tidal torques are included, the rotation  of close-in bodies can be captured in a resonant motion (Goldreich 1966; Goldreich \& Peale 1966; Celletti \& Chierchia 2008). The capture in a specific resonance depends on the eccentricity $e$ and on parameters related to tides and equatorial ellipticity ($Q$ and $C_{22}$, respectively). Hence, since $e$ is tidally damped during the inward migration, each specific capture should become unstable and the rotation can achieve another resonant state, which, at the same time, should result in a temporary trapping. When the orbit is circularized, the final evolution results in the synchronization between the orbital and rotational periods (i.e, the 1/1 spin-orbit resonance).

In this paper, we analysed the coupled spin-orbit evolution of Kepler-10\,\textbf{b}, GJ 3634\,\textbf{b} and 55 Cnc\,\textbf{e} short-period super-Earth planets. The numerical simulation of the exact equations of motion have shown that, depending on $C_{22}$, the rotation displays a number of resonance trappings for which $\Omega\simeq pn$, including the cases $p=5/2$, 2, 3/2, 4/3 and 1. Moreover, the results indicate that GJ 3634\,\textbf{b} can be an example of an exoplanet with a rotation state similar to Mercury's rotation ($p=3/2$), provided the body figure can support a permanent equatorial asymmetry as small as $C_{22}=10^{-5}$. 55 Cnc\,\textbf{e} can currently evolve under 5/2 or 3/2 trappings if $C_{22}\sim10^{-3}$ and under 2/1 if $C_{22}\sim10^{-4}$ or even if $C_{22}$ is as small as $2.45\times10^{-5}$. The rotation of 55 Cnc\,\textbf{e} may have a chaotic behavior for $C_{22}\sim10^{-3}$ (Callegari \& Rodr\'iguez 2012). However, we have shown that a scenario is possible in which the rotation evolves to synchronism through successive trappings in SOR. Indeed, the eccentricity damping due to tides provides the destabilization mechanism required to escape from one given capture and subsequently evolve to lower order resonances (see Correia \& Laskar 2004, 2009).

We have shown that different spin-states corresponding to various resonance captures can occur in extrasolar systems. Although the 1/1 spin-orbit  coupling is the final state of evolution, it is not reached in every case as present configuration. For current non-circular orbits, trappings in the 3/2 SOR or other resonance configurations can be sustained over long times. The latter would depend on the orbital eccentricity, the planet's $C_{22}$ value and its dissipation factor. Given that these parameters are not well-determined for terrestrial extrasolar planets, we note that the 1/1 coupling may not be the current rotation state of some planets. This would have consequences on the stellar insulation and therefore on the evolution of the atmosphere and surface of the planet.


When only the tidal torque is taken into account ($C_{22}=0$), the rotation evolves to the stationary tidal solution (STS). We have shown that under assumption of $C_{22}>0$, the evolution along spin-orbit resonance (SOR) leads to higher rates of orbital decay and circularization than the corresponding ones when the rotation evolves in the STS. Moreover, for a given $C_{22}$ value, the variation rates of the elements depend on the specific trapping in which the rotation is trapped in, and increases with $p$ (see Figure \ref{ae-zoom-k10}).

We have also calculated the averaged rates of variations of semi-major axis and eccentricity including the gravitational and tidal torques. Moreover, through the analytical calculation of the critical value of $C_{22}$ it is possible to predict the values of the eccentricity for which a given SOR would become unstable. The comparison with the direct numerical simulations have shown a good concordance for the analysed planets.

The importance of the $C_{22}$ value on the rotational history of the planet is illustrated in the example of Kepler-10\,\textbf{b}. It is worth noting that the planet rotation of Kepler-10\,\textbf{b} can evolve under the influence of the 2/1 SOR for longer time-scales (i.e., small $e$) provided the planet figure can support a permanent equatorial deformation characterized by $C_{22}\sim10^{-3}$. For smaller $C_{22}$, the 2/1 resonant state would become unstable for relatively large eccentricity (Figures \ref{c22} and \ref{rot-exce-k10}).

We stress that the results of the numerical simulations provide different possible spin-orbit evolutions depending on the considered initial values of elements and rotation. The $Q$ value and the specific tidal model would also play an important role in determining the rotation history of the planet, since the temporary trapping in SOR depends on the rate of eccentricity damping, which is $Q$-dependent. In fact, in a recent paper, Makarov (2012) has applied the model by Efroimsky (2012) and successfully explained the current resonant state of Mercury within the considered model.

Our results concerning the association of the critical $C_{22}$ value with the capture probability are not new in the context of averaged equations of the spin-orbit problem (see for instance Correia \& Laskar 2004, 2009). However, in this work, we are able to compare the results of the averaged equations with those resulting from the numerical simulations of the exact equations of motion.

A model of planet interior including the effect of viscous friction at the core-mantle boundary would provide quantitatively different results (see Peale 2005, and Correia \& Laskar 2009, for the case of Mercury). According to previous works the chances of resonant capture increase if the latter effect is taken into account (Goldreich \& Peale 1967; Peale \& Boss 1977). However, in order to highlight the contribution of $C_{22}$ to the orbital evolution, in this paper we adopt a simple model neglecting a possible layered structure of the planet.

In the present study we have assumed constant $C_{22}$ values. However, depending on the adjustment to hydrostatic equilibrium $C_{22}$ may be itself time-dependent. The time-scale for adjustment to the completely relaxed state depends on material parameters, e.g., rigidity and viscosity which are highly temperature dependent. Furthermore, it will depend on the spin-orbit coupling the planet is trapped in. For the application to terrestrial planets given here the constant, i.e. frozen, $C_{22}$ configuration is justified because adjustment of solid rocky planets to equilibrium usually occurs on geological time-scales. Therefore, the states described here do not necessarily represent hydrostatic configurations.


In this work we have considered a model of one interacting planet, despite the fact that Kepler-10\,\textbf{b} and 55 Cnc\,\textbf{e} belongs to multiple-planet systems with exterior (more massive) companions. Previous works have shown that the mutual secular interaction with the companion may affect the eccentricity evolution of the tidally evolved planet, increasing the circularization time-scale (Mardling 2007; Rodr\'{\i}guez et al. 2011a; Rodr\'{\i}guez et al. 2011b). Moreover, the planetary perturbations can produce large excursions of the innermost planet eccentricity, providing a mechanism for increasing or decreasing the capture probability of the spin-orbit resonances (Correia \& Laskar 2004, 2009). However, a numerical simulation of 55 Cnc system including the exterior companion 55 Cnc\,\textbf{b} has shown little differences with the single planet case. Indeed, the evolution of the inner orbit eccentricity is almost not affected by the mutual interaction between the planets and thus the rotation evolution of 55 Cnc\,\textbf{e} can be safely analysed within the one planet model. An investigation about the rotation of asymmetric close-in super-Earths including an exterior companion will be addressed in a future investigation.

\section*{Acknowledgments}
The authors acknowledge the support of this project by FAPESP (2009/16900-5 (A.R); 2006/58000-2 (N.C)) and CNPq. The authors also gratefully acknowledge the support of the Computation Centre of the University of S\~ao Paulo (LCCA-USP). We are grateful to the reviewer for his/her valuable comments.



\section*{Appendix A}

We show here the explicit expressions for the functions $H(p,e)$ and ${\cal{T}}(p,e)$. The exact form of $H(p,e)$ is given by

\begin{equation}\label{H}
H(p,e)=\frac{1}{\pi}\int_{0}^{\pi}\left(\frac{a}{r}\right)^3\mathrm{exp}(i\,2f)\mathrm{exp}(i\,2p\,l)dl.
\end{equation}
The expansion up to order six in $e$ is displayed in Table 2 for $p=1,3/2$ and 2.

\begin{tiny}
\begin{table}\label{tab2}
\begin{center}
\caption{\small Expansion of the function $H(p,e)$ given by equation (\ref{H}) for ${\cal O}(e^7)$ for the principal spin-orbit resonances.}
\begin{tabular}{c c c c c c c c}

\hline
\vspace{0.5cm}

   $p$ & $H(p,e)$ \\

\vspace{0.5cm}
 1 & 1 - $\frac{5}{2}e^2+\frac{13}{16}e^4-\frac{35}{288}e^6$ \\
\vspace{0.5cm}
 3/2 & $\frac{7}{2}e-\frac{123}{16}e^3+\frac{489}{128}e^5$  \\

 2 & $\frac{17}{2}e^2-\frac{115}{6}e^4+\frac{601}{48}e^6$  \\

  \hline
\end{tabular}
\end{center}
\end{table}
\end{tiny}

The exact expression for the function ${\cal{T}}(p,e)$ is given by (Hut 1981)

\begin{equation}\label{T}
{\cal{T}}(p,e)=(1-e^2)^{-6}\left[f_2(e)-(1-e^2)^{3/2}f_5(e)\,p\right],
\end{equation}
where
\begin{equation}\label{f2}
f_2(e)=1+\frac{15}{2}e^2+\frac{45}{8}e^4+\frac{5}{16}e^6
\end{equation}
and
\begin{equation}\label{f5}
f_5(e)=1+3e^2+\frac{3}{8}e^4.
\end{equation}




\begin{thebibliography}{}

\bibitem{1999Icar..142..219A} Agnor, C.~B., Canup R.~M., Levison H.~F., 1999, Icarus, 142, 219

\bibitem{2011ApJ...729...27B} Batalha N.~M.,
Borucki W.~J., Bryson S.~T., et al., 2011, ApJ, 729, 27

\bibitem{beutler05} Beutler G., 2005, Methods of Celestial Mechanics, vol. I, Springer, Berlin.

\bibitem{2011A&A...528A.111B} Bonfils X., Gillon M., Forveille T., et al., 2011, A\&A, 528, A111


\bibitem{calleg12} Callegari Jr. N., Rodr\'iguez A., 2012 (arXiv:1205.5704)

\bibitem{2007P&SS...55..889C} Celletti A., Froeschl\'{e} C., Lega E., 2007, PASP, 55, 889

\bibitem{2008CeMDA.101..159C} Celletti A., Chierchia L., 2008, Celestial Mechanics and Dynamical Astronomy, 101, 159

\bibitem{2006E&PSL.252..398C} Correia A.~C.~M., 2006, Earth and Planetary Science Letters, 252, 398 

\bibitem{Natur.429..848C} Correia A.~C.~M., Laskar J., 2004, Nature, 429, 848

\bibitem{Icar..201....1C} Correia A.~C.~M., Laskar J., 2009, Icarus, 201, 1

\bibitem{2012ApJ...751L..43C} Correia A.~C.~M., Laskar J., 2012, ApJL, 751, L43 

\bibitem{correia08} Correia A.~C.~M., Levrard B., Laskar J., 2008, A\&A, 488, L63

\bibitem{1962fcm..book.....D} Danby J., 1962, Fundamentals of celestial mechanics, New York: Macmillan


\bibitem{darwin1880}
Darwin G. H., 1880., Philos. Trans., 171, 713 (repr. Scientific Papers, Cambridge, Vol. II, 1908).

\bibitem{dobbs} Dobbs-Dixon I., Lin D.~N.~C., Mardling R.~A., 2004., AJ, 610, 464

\bibitem{2007Icar..192....1D} Dobrovolskis A.~R., 2007, Icarus, 192, 1 

\bibitem{2012CeMDA.112..283E} Efroimsky M., 2012, Celestial Mechanics and Dynamical Astronomy, 112, 283

\bibitem{efro2} Efroimsky M., Williams J.~G., 2009, CeMDA, 104, 257

\bibitem{everhart}
Everhart E., 1985, in Carusi A. and Valsecchi G. B., eds, Dynamics of comets: Their origin and Evolution. Reidel, Dordrecht, p. 185

\bibitem{sfm12} Ferraz-Mello S., 2012 (arXiv:1204.3957)

\bibitem{frh}
Ferraz-Mello S., Rodr\'iguez A., Hussmann H., 2008., CeMDA, 101, 171. Erratum: 2009., CeMDA, 104, 319

\bibitem{gi} Gillon M., Demory B.-O., Benneke B., et al., 2012, A\&A, 539, A28

\bibitem{G66}
Goldreich P., 1966, Astron. J., 71, 1

\bibitem{G666}
Goldreich P., Peale S., 1966, Astron. J., 71, 425

\bibitem{1967AJ.....72..662G} Goldreich P., Peale S., 1967, AJ, 72, 662 

\bibitem{1984Icar...58..186G} Greenberg R., Weidenschilling S.~J., 1984, Icarus, 58, 186 

\bibitem{2011A&A...528A..27H} Heller R., Leconte J., Barnes, R., 2011, A\&A, 528, A27 

\bibitem{hut81} Hut P., 1981, A\&A, 99, 126

\bibitem{2010Sci...327.1367I} Iess L., Rappaport
N.~J., Jacobson R.~A., et al., 2010, Science, 327, 1367

\bibitem{jackson09}
Jackson B., Barnes R., Greenberg R., 2009, ApJ, 698, 1357

\bibitem{kaula64} Kaula W.~M., 1964, Reviews of
Geophysics and Space Physics, 2, 661

\bibitem{2007ApJ...671.2082K} Kokubo E., Ida S., 2007, ApJ, 671, 2082

\bibitem{2012ApJ...752...73M} Makarov V.~V., 2012, ApJ, 
752, 73 

\bibitem{mard07b} Mardling R. A., 2007, MNRAS, 382, 1768

\bibitem{2011MNRAS.415.2275M} Michtchenko T.~A., Rodr{\'{\i}}guez A., 2011, MNRAS, 415, 2275 

\bibitem{mi79}
Mignard F., 1979, The Moon and the Planets, 20, 301

\bibitem{1999ssd..book.....M} Murray C.~D., Dermott S.~F., 1999, Solar System Dynamics, Cambridge

\bibitem{2005Icar..178....4P} Peale S.~J., 2005, Icarus,
178, 4

\bibitem{1965Natur.206Q1240P} Pettengill G.~H., Dyce R.~B., 1965, Nature, 206, 1240 

\bibitem{2010EAS....42..411R} Rodr{\'{\i}}guez A., Ferraz-Mello S., 2010, EAS Publications Series, 42, 411

\bibitem{2011MNRAS.415.2349R}
Rodr{\'{\i}}guez A., Ferraz-Mello S., Michtchenko T.~A., Beaug\'{e} C., Miloni O., 2011a, MNRAS, 415, 2349

\bibitem{2011CeMDA.111..161R}
Rodr{\'{\i}}guez A., Michtchenko T.~A., Miloni O., 2011b, Celestial Mechanics and Dynamical Astronomy, 111, 161

\bibitem{2004jpsm.book..281S} Schubert G.,
Anderson J.~D., Spohn T., McKinnon W.~B., 2004, Jupiter.~The Planet, Satellites and Magnetosphere, 281

\bibitem{2012Sci...336..214S} Smith D.~E., Zuber
M.~T., Phillips R.~J., et al., 2012, Science, 336, 214

\bibitem{1988Icar...74..172T} Tittemore W.~C., Wisdom J., 1988, Icarus, 74, 172


\bibitem{2011IAUS..276..181V} Valencia D., 2011, IAU
Symposium, 276, 181

\bibitem{2007ApJ...665.1413V} Valencia D.,
Sasselov D.~D., O'Connell R.~J., 2007, ApJ, 665, 1413

\bibitem{2012NatGe...5...18W} Wieczorek M.~A.,
Correia A.~C.~M., Le Feuvre M., Laskar J., Rambaux N., 2012, Nature Geoscience, 5, 18

\bibitem{2011ApJ...737L..18W} Winn J.~N., Matthews
J.~M., Dawson R.~I., et al., 2011, ApJL, 737, L18

\bibitem{2008Icar..193..637W} Wisdom J., 2008, Icarus, 193,
637





\end{thebibliography}
\end{document}